\documentclass[authoryear]{elsarticle}
\usepackage[latin1]{inputenc}
\usepackage{amsmath}
\usepackage{amsfonts}
\usepackage{amssymb}
\usepackage{natbib}

\begin{document}
\title{The Function of the Second Postulate in Special Relativity.}

\author{Alon Drory}
\ead{adrory@gmail.com}
\address{Afeka College of Engineering, 218 Bney-Efraim Street, Tel-Aviv, 69107 , Israel}

\begin{abstract}

Many authors noted that the principle of relativity, together with space-time symmetries, suffices to derive Lorentz-like coordinate transformations between inertial frames. These contain a free parameter, $k$, (equal to $c^{-2}$ in special relativity) which is usually claimed to be empirically determinable, so that special relativity does not need the postulate of constancy of the speed of light. I analyze this claim and find that all methods destined to measure $k$ fail without further assumptions, similar to the second postulate. Specifically, measuring $k$ requires a signal that travels identically in opposite directions (this is unrelated to the conventionality of synchronization, as the one-postulate program implicitly selects the standard synchronization convention). Positing such a property about light is logically weaker than Einstein's second postulate but suffices to recover special relativity in full.
\end{abstract}


\maketitle

\section{Introduction}
\label{intro}

In his famous 1905 paper, Albert Einstein based the theory of special relativity (SR) on two principles, or postulates:

\begin{quote}
1. [Principle of relativity] The laws governing the changes of the state of any physical system do not depend on which one of two coordinate systems in uniform translational motion relative to each other these changes are referred to.

2. [Principle of the constancy of the speed of light] Each ray of light moves in the coordinate system ``at rest'' with the definite velocity V independent of whether this ray of light is emitted by a body at rest or in motion. \citep[p. 895, English translation from The Collected Papers, Vol. 2, 1989, p.143]{einstein1}
\end{quote}
Homogeneity of space and time and isotropy of space, implicit in Einstein's paper, were later explicitly added to the list of postulates.

From as early as 1910 with \citet{ignatowski1,ignatowski2}, and \citet{frank}, many researchers questioned the necessity of the second postulate  \citep[e.g.,][ among others]{weinstock,mitvalsky,levy-leblond,srivas,mermin,schwartz,sen,pal}. They claimed that, up to an empirically determinable universal constant, the coordinate transformations of special relativity can be derived without using any information about light, and notably without any postulate about the invariance of its speed.

The form of the coordinate transformations between two inertial frames of reference $S$ and $S'$, turns out to be severely restricted by the following three postulates:

P1. Homogeneity of space and time: The laws of physics are invariant under a translation of the origin of coordinates of space and time. 

P2. Isotropy of space: The laws of physics are invariant under rotations of the axes in which they are described.

P3. Principle of relativity: The laws of physics are invariant under a transformation between two reference frames in uniform relative motion.\footnote{Strictly speaking, one should add to these postulates the law of inertia (Newton's first law). However, to obtain this from the principle of relativity one must merely assume that a body at rest remains at rest if it is an isolated body (under no external influence). This is usually taken to be a basic principle of sufficient cause and is thus seldom listed as a separate assumption. The principle of relativity then implies that an isolated body in uniform motion must also persist in this state, since a body at rest in $S$ appears to a moving observer $S'$ to be in uniform motion.}

From postulates P1-P3 alone, one can then derive what I shall term ``generalized Lorentz transformations'', to distinguish them from the standard Lorentz transformations, which contain the speed of light. These are \citep[for the derivation see, e.g.,][]{levy-leblond,drory}:

\begin{eqnarray}
\label{final}
x' = \dfrac{\left( x - vt \right)}{\sqrt{1 - kv^2}} \nonumber \\
t' = \dfrac{\left( t - kvx \right)}{\sqrt{1 - kv^2}}
\end{eqnarray}

As usual, $x', t'$ refer to the coordinates of some event observed by an observer $S'$. $S'$ moves at a constant velocity $v$ with respect to a second observer $S$, who ascribes to the same event  the coordinates $x, t$. The parameter $k$ is a universal constant, such that $k=0$ corresponds to the Galilean transformations.

From these we obtain immediately the rule of addition of velocities:
\begin{equation}
\label{veladd}
w = \dfrac{v + u}{1 + kuv} .
\end{equation}
Here $u$ is the velocity of a body moving in the $x$-direction, as measured by $S$. $w$ is the same body's velocity in the frame $S'$. 

It is hardly obvious that the relativity principle and the space-time symmetries by themselves restrict the possible coordinate transformations to just two types: Galilean and Lorentz-like. \citet{pal} justly stressed this point in his own derivation. These transformations are thus of great interest but their significance is not entirely clear. An early critic was \citet[p.11]{pauli} who wrote (alas without detailing his argument) that although one could derive the form of the transformations Eqs.(\ref{final}), the same was not true of their physical content.

In a previous work I have argued that as a physical theory, special relativity must be explicitly distinguished from Newtonian mechanics, since these two theories imply different ontologies, different procedures for reproducing experiments, etc...\citep{drory}. This means that the cases $k = 0$ and $k \neq 0$ represent two different theories, each distinguished from the other by an additional postulate regarding the value of $k$. While that value must of course be determined empirically, the distinction between the vanishing and non-vanishing cases is so fundamental that it must be included among the postulates of the theory before we can speak of the resulting structure as a well defined physical theory at all.

In the present work I concentrate on the value of the undetermined universal parameter $k$, which is equal, in standard SR, to $\dfrac{1}{c^2}$. A.M. Srivastava's attitude seems quite common when he notes:

\begin{quote}
As we know, the experiments show that [$k^{-1/2}$] has a finite value which is equal to the value of the speed of light. \citep[p. 505]{srivas}
\end{quote}

Similarly, \citet{mermin} explains:

\begin{quote}
From this point of view, experiments establishing the constancy of the speed of light are only significant because they determine the numerical value of the parameter k.\citep[p. 119]{mermin}
\end{quote} 

Most authors on the subject take such a measurement to be a trivial operation (N David Mermin being a notable exception and I shall analyze his insights in detail). In their view, the second postulate is nothing more than a report of the experimental value of $k^{-1/2}$. To these authors, it is utterly contingent that this value turns out to be the speed of light, and neither light itself, nor any other signal propagating at $k^{-1/2}$ plays any role in these derivations. Mermin goes even further in declaring:

\begin{quote}
It is not, however, necessary for there to be phenomena propagating at the invariant speed to reveal the value of k.\citep[p. 123]{mermin}
\end{quote} 

My aim here is to show that this claim is false. Experimentally measuring $k$ runs into serious difficulties if we do not use something akin to the second postulate. In particular I will analyze the above claim by Mermin and show where it goes wrong (section\ref{sec:mermin} and \ref{sec:mermin2}).

Other methods of obtaining $k$ also fail on the same grounds, namely because we do need some phenomenon that propagates isotropically irrespective of the speed of its source (sections \ref{sec:doppler} and \ref{klength}). Furthermore, the required signal must in fact propagate at the invariant speed, contrary to Mermin's claim (section \ref{sec:synchro}). There is no compelling reason why this signal should be light, however, other than the fact that it is so in our world. In this sense, and only in this sense, light is not necessary to special relativity. But some form of second postulate, declaring an invariance property of some signal is necessary in principle.

Section \ref{sec:weaker} then presents some general considerations why special relativistic theories need a postulate about the properties of specific signals. I show there that one can recover the standard theory of special relativity by adding to the fundamental symmetries a postulate on the isotropy of propagation of light. Although logically weaker than Einstein's second postulate, it is sufficient to derive the invariance of the speed of light. The paper then closes with a brief summary of the main points.

\section{Aims and Tools}
\label{sec:rules}

The one-postulate program is this: using only the three postulates P1-P3 (the relativity principle and spacetime symmetries), derive the special relativistic kinematics up to a numerical constant that is determinable in principle \textit{without invoking further physical knowledge}.

The postulates \textit{must} include all the physical knowledge utilized, otherwise the principle of relativity itself could be omitted either because it is experimentally well established or because it is merely an integral part of mechanics. Similarly, all knowledge of electromagnetism must be put aside else the second postulate is sneaked in through the back door.

Deriving special relativity (rather than some other, formally similar, theory) requires that we can measure $k$ to arbitrary precision in principle. To this end, we use gedanken experiments, where we can ignore technical difficulties and experimental errors, but no essential effects and we cannot break physical laws. As we shall see, it can be difficult to distinguish a technical difficulty from a fundamental effect and care must be exercised here. 

We must remember that we are concerned with the logical structure of the theory, not its correctness. Thus we must ignore any limitations on the validity of the postulates, e.g., from general relativity or quantum mechanics, since they would just as well invalidate the standard derivation of special relativity. In the present analysis we take the postulates to be true, even if we happen to know that in fact they are not or that they have a limited domain of validity.

Finally, note that $k \neq 0$ and $k = 0$ suppose different preconditions of experiments. If $k \neq 0$, durations and lengths are not absolute. Clocks and rods will be affected by their motion and this must be taken into account when devising experimental procedures. We do not presuppose whether $k$ vanishes or not, whether it is large or small. But any method of measurement of $k$ must cover all options, which means that we must devise it \textit{as if} $k \neq 0$ and large enough to influence our instruments. This does not preclude that we should eventually find that $k = 0$ or close to it. Any procedure consistent with the assumption that $k \neq 0$ will also be consistent with the possibility that $k = 0$, but the reverse is not true. Clock synchronization, for example, can be done by arbitrarily quick transport if $k = 0$ but obviously not if $ k \neq 0$. On the other hand, the Einstein synchronization procedure is valid even if $k = 0$.

In the following, therefore, we must set up all experiments as if $k \neq 0$, although in fact we make no assumptions as regards the actual value of $k$, which may be zero, low or high.

With these caveats in mind, let us see how we could measure $k$.

\section{Finding $k$ from the Velocity Addition Formula}
\label{sec:mermin}

N. David Mermin (\citeyear{mermin}) devised a highly ingenious method of measuring $k$ from the velocity addition formula, Eq.(\ref{veladd}). Mermin's original exposition contains many extraneous details and I have adapted it here to the problem at hand.

If an observer A wants to measure the velocity of an object B, he must choose a set of units. Instead of length and time, A can use units of velocity by comparing the velocity of B to that of a unit runner (a ``hare'' in Mermin's terminology). We assume that A has an object that moves at a velocity $u_A$ on a straight track for a certain length L, then stop and run back at the same speed in the opposite direction. Let us also assume that B moves more slowly than the unit runner, so that the unit runner reaches the end of its track and doubles back before B has reached the end of the same track. Finally, let us assume that both B and the unit runner start at $t=0$ from the beginning of the unit runner track.

Let $t_1$ be the time at which the unit runner reaches the end of its track:

\begin{equation}
\label{mermin1}
u_A \cdot t_1 = L
\end{equation}

The runner turns around when the body B is still moving in its original direction (towards the end of the runner's track). Let $t_2$ be the time from the runner turning around until it meets the body B again. Then,

\begin{equation}
\label{mermin2}
v_{BA}(t_1 + t_ 2) + u_A t_2 = L   ,
\end{equation}
where $v_{BA}$ is the velocity of the object B in the frame of reference A. The unit runner meets B a distance $L_0$ from the track's end, so that

\begin{equation}
\label{mermin3}
L_0 = u_A t_2
\end{equation}

The ratio $\dfrac{L_0}{L}$ is a frame invariant quantity. For example, one can imagine the distance L being broken into N separate length units numbered from 1 to N. All observers will agree on which number $n$ represents the particular unit where the meeting happens, because the segment $n$ could light up when the meeting happens with all the other segments remaining turned off. As $N$ grows to infinity, the ratio $\dfrac{n}{N}$ approaches the value $\dfrac{L_0}{L}$ , which must also be frame invariant, therefore. Isolating $t_1$ and $t_2$ from Eqs.(\ref{mermin1}) and (\ref{mermin3}) then substituting these into Eq.(\ref{mermin2}), we obtain:

\begin{equation}
\label{ratio}
\dfrac{v_{BA}}{u_A} = \dfrac{L - L_0}{L + L_0} = \dfrac{1 - \left(\dfrac{L_0}{L}\right)}{1 + \left(\dfrac{L_0}{L}\right)}
\end{equation}

This yields the velocity of B in terms of the unit runner velocity, $u_A$, without the need for clocks or rulers.

To measure $k$, we now proceed as follows: let A and B be two frames moving in parallel in the same direction. Each has a unit runner moving with respect to its own frame at a velocity $u_A$ and $u_B$ respectively. $u_A$ and $u_B$ need not be identical, since only velocity ratios enter the final expression. Without loss of generality, let us assume that $u_A > u_B$. A third object, C, moves in the direction of motion of B (relative to A). Using the unit runner method, A measures the velocity of C and B and finds the two ratios 

\begin{eqnarray}
r_B = \dfrac{v_{BA}}{u_A} \nonumber \\
r_C = \dfrac{v_{CA}}{u_A}
\end{eqnarray}
Similarly, B measures the velocities of A and C with his own unit runner and finds the ratios

\begin{eqnarray}
s_A = \dfrac{v_{AB}}{u_B} \nonumber \\
s_C = \dfrac{v_{CB}}{u_B}
\end{eqnarray}

We can now use the velocity addition rule, Eq.(\ref{veladd}), to obtain:

\begin{equation}
\label{vca}
v_{CA} = \dfrac{v_{CB} + v_{BA}}{1 + kv_{CB}v_{BA}} = v_{BA}\left[\dfrac{1 + \left(\dfrac{v_{CB}}{v_{BA}}\right)}{1 + k \left(\dfrac{v_{CB}}{v_{BA}}\right)v_{BA}^2}\right]
\end{equation}

Finally, we note that $v_{BA} = - v_{AB}$. This lets us rewrite

\begin{equation}
\dfrac{v_{CB}}{v_{BA}} = - \dfrac{v_{CB}}{v_{AB}} = - \dfrac{s_C}{s_A}
\end{equation}

Eq.(\ref{vca}) then becomes

\begin{equation}
\label{rc}
\dfrac{r_C}{r_B} = \dfrac{1 - \left(\dfrac{s_C}{s_A}\right)}{1 - ku_A^2 \left(\dfrac{s_C}{s_A}\right)r_B^2}
\end{equation}

From this we finally obtain, 

\begin{equation}
\label{kmermin}
ku_A^2 = \dfrac{1}{r_Br_C} \left[ 1 - \dfrac{s_A}{s_C}\left( 1 - \dfrac{r_C}{r_B}\right) \right]
\end{equation}

This gives the invariant velocity $k^{-1/2}$ in terms of the velocity $u_A$ of the unit runner of the frame A. This value is expressed entirely through dimensionless frame-invariant ratios, for which we need neither clocks nor rulers. 

It seems that we have succeeded, then, but for one difficulty, which, as we shall see, recurs over and over again in all the attempts to measure $k$. The measurement requires ``\textit{a mechanism of propulsion that couples each hare [unit runner] to its track in the same way for the right-left and left-right parts of its race}'' \citep[p. 124]{mermin}.

But how can we be assured that the runners move identically on the left-right and right-left parts of their track? At first sight, this seems easy: build two identical motors and rails. To make sure that they are identical, run the two runners in the same direction and confirm that they are in identical positions at every instant (we can assess this visually without the need for clocks). Then we rotate the rails and motors of one runner by $180^0$ and use that for the second leg of the runner's journey.

But how do we know that the rotation hasn't affected the working of the motors? Compare the situation with that of another mechanical device, viz., a clock. We know that rotating a clock does in fact alter its ticking rate, therefore, rotating mechanical devices may, in general,  influence their working. What guarantees that the motor's workings are not analogously affected by the rotation? Barring a complete description of every detail and structure of the transportation system, and an equally complete understanding of every physical effect involved, we cannot be certain of this. Even in Newtonian physics, rotating a motor is a bad way to ensure identical motion in two different directions because accelerations produce forces that might, in principle, bend, twist or otherwise change some part of the mechanism. To be sure, this might be unlikely, or negligible, but in gedanken settings, where absolute precision is obtainable, we must be careful, because we want to determine $k$ in principle, not merely find a ``good enough'' method for practical purposes. Remember that the proponents of the one-postulate approach must devise a method of measuring $k$ \textit{in principle}, i.e., to arbitrary precision. Contingent trial and error methods are not what we seek, no matter how useful they are in actuality and history.

Luckily, we seem to have an easy solution to the dilemma. Instead of physically rotating the motors, we can rotate their \textit{plans} and build the moving mechanism from the rotated design. Spatial isotropy would then seemingly ensure that the rotated structure functions identically to the original one.

But does spatial isotropy actually ensure this result?

\section{Two meanings of isotropy}
\label{isotropy}

The term ``isotropy", and more generally the notion of (space-time) symmetries, covers two related but nevertheless distinct concepts. The first is a general principle, that the form of the laws of physics have certain invariance properties. For isotropy, this means that the laws show no preferred directionality (rotational invariance). We can call this ``nomological symmetry'' or specifically ``nomological isotropy". The second possible meaning of ``isotropy" is the absence of preferred directionality in a specific experiment performed under well-defined conditions. In this case isotropy refers to invariance under an actual physical rotation of the system. Such a property may be termed ``systemic isotropy" or more generally ``systemic symmetry".
 
\subsection{Nomological symmetries}

Nomological symmetry is an abstract, mathematical property of the laws. It states that the expression of a law, or a number of laws, is covariant with respect to some transformation, e.g., of the coordinate system. Thus, rotational invariance of the fundamental laws of mechanics means that these laws retain their mathematical form under a rotation of the axes, after all the vector components have been recalculated in the rotated frame. 

The relevant transformation here is a rotation of the coordinate system, not a physical rotation of the system. Such a mathematical transformation is of course equivalent to a physical rotation of the system in the opposite direction while keeping the axes unchanged, but the definition of the system is then crucial. A rotation of the axes is a global transformation, not a local one. It is thus only equivalent to a physical rotation of the entire universe in the direction opposite the rotation of the axes. We expect that rotating the entire universe is geometrically meaningless (but not necessarily dynamically so, as suggested by Mach's principle), because we believe that the orientation of coordinate systems is arbitrary and should carry no meaningful content. That is the content of nomological isotropy.

This definition might seem too confining. Noether's theorem, for example, applies to any Lagrangian that is invariant under a continuous symmetry. Nowhere does it require that this Lagrangian should describe the whole universe. But this is only because here as with all their uses, nomological symmetries apply to abstract systems, not to actual objects. In Noether's theorem, a rotation still means a mathematical transformation of the coordinates, not a physical rotation of the system or of part of it. By limiting ourselves to a specific Lagrangian, we are in effect creating a toy universe which contains only the terms referred to in this Lagrangian. We are thus imagining that the rest of the universe has been erased, and that the whole of actuality is described by the terms we include in the Lagrangian. The equivalent physical rotation must still apply to the entirety of our fictional, toy universe.

Whether such a fiction is useful or not is a different matter. Clearly, when we discuss nomological isotropy in this way, we are imagining that every \textit{relevant} factor has been rotated, and relevance usually depends on the desired precision.

Consider for example ballistic motion. The general laws of dynamics are isotropic, so we might expect that it exhibits invariance under horizontal rotations (clearly, full three dimensional isotropy is broken by the preferred direction of gravity). What we mean by that depends on the level of description that we use, however, i.e., how rich we wish to make our toy universe. That, in turn, depends on the required degree of precision.

The simplest level will only have gravity. In this approximation the laws are horizontally isotropic, so that objects thrown from identical height at identical horizontal speeds will land identical distances away from their initial point, irrespective of the horizontal direction of the throw. 

In actuality, however, objects thrown northwards and eastwards will travel different distances because of the Coriolis force due to the earth's daily rotation. A more refined description will include the earth's angular velocity vector, which selects the direction of the Earth's spinning axis as special. Nomological isotropy is now conditional on the understanding that when we rotate the axes, the angular velocity vector is also rotated in the required way. Thus, nomological isotropy here is equivalent to the physical rotation of a system that includes our planet. In effect, the direction of projection with respect to the geographic cardinal points remains invariant; east remains east under the rotation, although it would point to a different direction with respect to a coordinate system exterior to earth.

For higher precision, however, we might need to consider the rotation of the earth around the sun. Objects thrown from earth into space, for example, might feel a Coriolis force due to the planet's yearly rotation. Nomological isotropy is then valid only if rotating the axes includes rotating the angular velocity vector of the earth around the sun. The physical rotation equivalent to this mathematical transformation must extend to the entire solar system (or at least include the sun). As more and more precision is required, we might need to include our arm of the galaxy, the galaxy itself, the local group and so on. While these are usually irrelevant in practice, for questions of principle in gedanken settings the isotropy of the laws can only be equivalent to a physical rotation of the entire universe. 

\subsection{Systemic Isotropy}

Nomological isotropy is concerned with abstract properties of the mathematical descriptions of systems. In empirical situations, however, we often use the term ``isotropy'' differently. 

Suppose that we are performing a measurement with an apparatus containing an elongated arm of some sort, and we wonder whether the result exhibits any directionality. To check this, we physically rotate our apparatus, including the object on which the measurement is performed, without changing any other parameters. If we obtain the same result as before, our system exhibits a kind of rotational invariance that we may call ``systemic isotropy", or in general ``systemic symmetry".

The essential difference between this and nomological isotropy is that here, the operation under which the system is invariant is a physical rotation of the local system investigated, not an abstract rotation of the axes. Because physical rotations cannot be performed on the entire universe, they necessarily apply to a small, local set of objects, of which we have physical control. The rotational invariance of these systems then depends on the degree to which they are independent on the external world or influenced by it. 

Let us return to the simple example of ballistic motion. We cannot as a matter of fact rotate the earth or the solar system. Physical rotations are limited to the direction in which the object is thrown and possibly some local conditions (e.g., the experiment is performed in a sealed vacuum chamber that may also be rotated). If our experiment is sensitive enough to pick up the effects of the Coriolis force, we shall not have systemic isotropy. 

In applying nomological isotropy, we can formally rotate all relevant vectors, including the angular velocity of earth. But physically, we can only rotate some of the vectors, e.g., the initial velocity of the object. The planet's angular velocity is unavailable to manipulations. Systemic isotropy concerns a physical transformation performed on the system, or a physical symmetry that the system possesses, not a mathematical manipulation of the axes used to describe this system. Thus, systemic isotropy is entirely an empirical matter to be decided by comparing the results of experiments that differ only in the physical orientation of the apparatuses used to perform them. 

Systemic and nomological symmetries are completely separate and independent properties, neither of which entails the other. Systemic symmetry is a contingent property and we have many cases of events that exhibit no systemic symmetry, although the general laws of physics that determine them are nomologically invariant.  

Conversely, it is possible that although a certain class of phenomena is devoid of nomological symmetry, a specific system, through, e.g., some accidental cancellation of effects, should exhibit this symmetry systemically. For example, there is no law to the effect that every object should have an existing mirror image. The number of right-handed corkscrews is vastly superior to the number of left-handed ones. But a system of socks does usually exhibit such a symmetry so that at least at the manufacturing level, every sock of a given pattern usually possesses a mirror-symmetry twin.

The structure of special relativity itself reveals this twin aspect, for the postulate of isotropy is clearly meant to be nomological. In the derivation of the Lorentz transformations (whether generalized or not) it is applied to the axes themselves, or equivalently, to the frames of reference, where these are supposed to extend over the entire universe. Again, whether such a supposition holds in practice is a different matter. General relativity restricts the postulates of special relativity to local frames only. This represents a limitation on the empirical adequacy of special relativity in our world, not a limitation on the use of these postulates within the formal structure of the theory. 

On the other hand, the postulate of the constancy of the speed of light decrees (among other things) a certain systemic isotropy, namely, that all light signals propagate at the same speed in all directions, for all states of the emitting body. Obviously, the nomological isotropy of the laws that is assumed in the derivation of the Lorentz transformations is insufficient to guarantee this particular property of light. Even when isotropy is allied with the principle of relativity and space-time homogeneity, we only obtain that \textit{some} velocity $\left( \dfrac{1}{k^{1/2}} \right)$ must exhibit such invariance, but this limit velocity need not be related to light (though of course it turns out be so in our world). 

Conversely, the assumption that the speed of light possesses this systemic isotropy is not in itself enough to guarantee that the transformations themselves will exhibit nomological isotropy. Indeed, I previously obtained anisotropic transformations that nevertheless embody the isotropic propagation of light signals \citep{drory}. These transformations are not nomologically isotropic, but light propagation remains systemically isotropic under them.

Here again, it is important to note that the systemic isotropy of light is postulated, and that the question of its empirical adequacy is not relevant to the present discussion. Thus, general relativity limits this systemic symmetry to flat regions of space-time, but this only means that the second postulate, and hence special relativity itself, has limited applicability. The question under discussion here, however, assumes that the postulates are correct and seeks to determine the logical structure of their relations.

Though it might appear trivial at this point, the distinction between these two meanings of ``isotropy'' will be helpful in the remainder of the present analysis. Empirically determining $k$ turns out to essentially involve questions of isotropy. When we consider specific effects or specific apparatuses meant to measure $k$, the required isotropy is always systemic; it is then crucial to remember that such a symmetry is not guaranteed by the nomological symmetry postulated in the derivation of the generalized Lorentz transformations. Instead, if a suggested empirical procedure for measuring $k$ depends on the property of isotropy (as several will prove to in the next sections), it will have to be analyzed on a per case basis. In no case is the isotropy referred to in such cases, which is clearly systemic, derivable from the nomological isotropy postulated in the theory.

\section{Mermin again}
\label{sec:mermin2}

Let us now return to Mermin's procedure. We wish to ensure that two runners are moved at identical speeds but in opposite directions by building identical moving mechanisms and then inverting one of them. But what type of isotropy are we invoking here when we assume that the motor built according to rotated blueprints will perform the function we seek? 

At first sight, we are using nomological isotropy since we are merely rotating the axes of the coordinate system. This seems to solve the problem until we reflect that when we speak of the moving mechanism, we must consider everything that can influence the motion of the runner. That must include every object that \textit{might} exert a force on the runner. Rotating the motor and rails is not enough. If we want to include every \textit{possible} influence, we must rotate the whole universe. In principle, the gravitational influence of a single star might affect the motions of the runners by speeding up the velocity of the runner that moves towards the star, while braking the one that moves away from the star. Still more factors, such as local electromagnetic fields and other influences, must also be considered.

We might tend to dismiss such effects as negligible, but unfortunately, we cannot make such a claim without knowing beforehand something about the value of $k$. If $k$ is large enough, the measurement might be influenced by extremely minute differences in the configuration of the external world. Recall that we are not privy to any details about the value of $k$, other than the fundamental assumption that we should treat it as non-zero. Nothing about its actual value can be assumed \textit{beforehand}. Even if it were, in a gedanken setting precision is absolute, so even small influences must be taken into account.

One may wonder why no such problem arises in standard special relativity. The Einstein synchronization procedure is apparently vulnerable to a similar critique because gravitational influences affect the speed of light. The anisotropy of the world itself should impact the use of light signals to synchronize clocks, therefore \footnote{It is very important to distinguish this \textit{physical} anisotropy in the speed of light in some directions because of local interactions from the \textit{conventional} anisotropy of the one-way speed of light that is brought up in discussions of the conventionality of simultaneity. Section \ref{convent} discusses this issue further.}. But in special relativity, which is our present concern, light is \textit{postulated} to move at a universally constant speed. Whether the second postulate is true in our world is an empirical matter, but in the derivation of special relativity, it is posited as an axiom. In other words, the constancy of the speed of light is an assumption that belongs to the logical structure of the theory; its empirical verification belongs to realm of the theory's confirmation.

Therefore, that the speed of light in a vacuum is not influenced by any external influence is \textit{explicitly} guaranteed, \textit{in the context of standard special relativity} by the postulates themselves. The situation is completely different in the case of the mechanically moving runners in Mermin's procedure. No additional postulate \textit{logically} ``protects" the mechanisms from outside influences. The problem here is not that we assume something about the workings of the transportation mechanisms and then face the question of its empirical adequacy. Rather, the problem is that no additional assumptions are made at all. Nothing allows us to logically expect that rotated mechanisms will work identically if we do not rotate everything else around them. The reason is that it is unclear what additional assumptions will ensure that the motion of a runner are identical in opposite directions. 

It is not enough, for example, to postulate that gravitational influences are negligible. That would make the situation similar to special relativity, where we postulate a condition that we know to be violated, e.g., in the presence of strong gravitational fields. But here gravitation is just one of infinitely many possible influences. We need a complete theory of mechanics in order to know what is relevant and what is not in constructing the rotated machines. Thus, the kinematics of the theory, which is where we find the parameter $k$, depends on its dynamics, quite unlike what happens in special relativity where a single assumption covers every contingency. 

Furthermore, we expect that the dynamics should be founded on the kinematics rather than the other way around. The only non-dynamic assumption that will make Mermin's procedure work is if the structure of the entire universe is isotropic down to local details. While logically possible, such a postulate is problematic for being obviously violated in our world. Furthermore, adopting an adequate ``protecting'' postulate would lose us any logical advantage we supposedly gain from working with the principle of relativity alone.

Actually, the situation is worse,because Mermin's method requires that the left-to-right and right-to-left motions not only be identical,  but also uniform. Eq.(\ref{mermin2}) only holds for constant speeds. If the speeds are varying, we need to know their precise dependence on time in order to find the relation between $\dfrac{L_0}{L}$ and $k$. It is unclear then that we could extract $k$ from merely the ratio $\dfrac{L_0}{L}$, and we would likely need more parameters. 

The problem is how we can know that the unit runners are moving uniformly. Mermin simply assumes that we can create a mechanism that does that. This is far from evident, however. To devise such a mechanism theoretically requires a complete theory of mechanics so that we know what forces are at work and how they influence the runner's motion. Since we are in a gedanken setting, we are not allowed to dismiss any influence as negligible. We are imagining absolute precision, so once more even a distant star's gravitational influence, or a local charge's electric field have to be taken into account. But we cannot know all the influences, nor do we know exactly what the dynamical laws that determine these influences are. All this is beyond the scope of special relativity, and to require such complete knowledge in order to make sense of the theory is unreasonable.

The alternative is to adjust the mechanism by trial and error until it actually causes the runner to move uniformly. But then we need a method of determining that this goal has been achieved. The most direct way would be to check that for any set of equal time intervals $\Delta t$, the runner advances equal distances, $\Delta x$. But this requires a network of synchronized clocks that allows us to determine the position of the runner at equal time intervals $\Delta t$. Section \ref{sec:synchro} will explore this problem in detail.

In the next section, however, I explore a method that bypasses the need for synchronized clocks. As we shall see, this method also fails.

\section{Finding $k$ from the Doppler Shift}
\label{sec:doppler}

Since $k^{-1/2}$ has the units of velocity, any determination of its value must be relative, i.e., we compare $k^{-1/2}$ to some other velocity chosen to be a standard unit. In the previous section, we measured $ku_A^2$, where $u_A$ is the velocity of a mechanical runner. Another option is to compare $k^{-1/2}$ with the speed of some standard non-material signal and using the Doppler effect. 

Suppose that we wish to measure the velocity $v$ of an object moving in the positive $x$ direction towards an observer S. The body emits a periodic signal that also travels in the positive $x$-direction. Let the frame S' be the rest frame of the body, in which the (rest) frequency of the emitted signal is $f_0$, and its period is $T_0 = \dfrac{1}{f_0}$.

In the S frame, on the other hand, the body emits a signal with a period

\begin{equation}
T = \dfrac{T_0}{\sqrt{1 - kv^2}}
\end{equation}
in accordance with the generalized time dilation rule. 

Let the velocity of the signal in the S frame be $c$ (we assume nothing regarding its velocity in the S' frame). At a certain moment, the moving body emits a wave crest, for example. By the time the emitter sends the next crest, the first one has moved a distance $cT = \lambda$ towards the observer. But the emitter has also advanced, by a distance $v T$. Thus, the distance between one crest and the next is only 

\begin{equation}
\lambda_{obs} = \left( c - v \right) T 
\end{equation}

The period observed by S is the time between these two crests, which is 

\begin{equation}
T_{obs} = \dfrac{\lambda_{obs}}{c} = \left[1 - \dfrac{v}{c} \right] \dfrac{T_0}{\sqrt{1 - kv^2}}
\end{equation}

The observed frequency is therefore

\begin{equation}
\label{gdoppler}
f_{obs} = \dfrac{1}{T_{obs}} =  \dfrac{\left[1 - \dfrac{v}{c} \right]}{\sqrt{1 - (kc^2)\frac{v^2}{c^2}}} f_0
\end{equation}

In this equation, the generalized Doppler formula, all the velocities, including the invariant velocity $k^{-1/2}$, are expressed relative to the signal velocity $c$ that can be chosen as a unit velocity.

To measure the ratio $\dfrac{f_{obs}}{f_0}$ (or equivalently, $\dfrac{T_{obs}}{T_0}$), the observer S needs only one clock by which he measures the frequencies emitted, once by the body in motion and once by an identical signal emitter at rest. There is no need to synchronize this clock with another. The measured ratio $\dfrac{f_{obs}}{f_0}$ contains the desired parameter $kc^2$. 

Unfortunately, it also contains the emitter velocity ratio, $\dfrac{v}{c}$. Because $c$ has already been selected as the unit velocity, $\dfrac{v}{c}$ can no longer be chosen as unity and must be determined as well. Thus, Eq.(\ref{gdoppler}) contains two unknowns and only one measured quantity, the frequency ratio, to determine them. We cannot extract the desired parameter $kc^2$ unless we have another equation involving the same unknowns (and only them). 

Suppose we reverse the direction of motion of the object, so that $v \rightarrow -v$. The emitter now moves away from the observer, and from Eq.(\ref{gdoppler}) we have now:

\begin{equation}
\label{gdoppler2}
f_{obs,2} = \dfrac{\left[1 + \dfrac{v}{c} \right]}{\sqrt{1 - (kc^2)\frac{v^2}{c^2}}} f_0
\end{equation}

Together with Eq.(\ref{gdoppler}) we could now extract the value of $kc^2$. But we immediately encounter the problem we faced with Mermin's procedure: how can we make sure that the object will move with a velocity $-v$? This is like reversing the motion of a unit runner and we run into the same difficulties.

A seemingly better choice is to keep the body moving in the same direction, but repositioning the observer behind it. The effect is the same, namely, that the signal is now emitted from a body moving away from the observer, but no mechanical inversion took place, so the body keeps moving as before. Unfortunately, the signal does not. We would like to claim that this situation is related to the previous one (the body moving towards the observer) by performing the simple replacement $c \rightarrow -c$. That would yield once again Eq.(\ref{gdoppler2}). But we cannot, because nothing guarantees that the signal's speed when the emitter moves away from the observer is the same as when the object moves towards him. Indeed, if we use sound waves and the observer is not in the rest frame of the propagation medium (e.g., air), we know that these two speeds will be different. What we must do is then to assume one speed for motion towards the observer, $c_+$, and another for motion away from the observer, $c_-$. Then the two equations we obtain for the observed frequency ratios are:

\begin{eqnarray}
\label{isotrodop}
\dfrac{f_+}{f_0} =  \dfrac{\left[1 - \dfrac{v}{c_+} \right]}{\sqrt{1 - (kc_+^2)\dfrac{v^2}{c_+^2}}} \nonumber \\
\dfrac{f_-}{f_0} =  \dfrac{\left[1 + \dfrac{v}{c_-} \right]}{\sqrt{1 - (kc_-^2)\dfrac{v^2}{c_-^2}}}
\end{eqnarray}

This renders the method useless, of course. While the two ratios on the left hand sides of the equations can be measured, we have four unknowns, $\dfrac{v}{c_+}$, $\dfrac{v}{c_-}$, $kc_+^2$ and $kc_-^2$, which still leaves us unable to determine $k$. Of course, Einstein's second postulate solves this problem by \textit{postulating} that $c_+ = c_-$, which reduces the number of unknowns to the required two. Yet this is precisely the assumption that we seemingly want to avoid. Once again, we are at a dead end.

\section{Finding k from the transformations}
\label{klength}

Next, we could try to determine $k$ from the generalized Lorentz transformations themselves. Two options present themselves. 

The first is to use the contraction of lengths. As in SR, we assume a rod with rest length $L_0$  placed in a S'-reference frame moving at velocity $V$ with respect to an observer S. S simultaneously locates the rod's end points and measure its length to be

\begin{equation}
\label{contraction}
L_0 = \dfrac{L - vt}{\sqrt{1 - kV^2}} - \dfrac{0 - vt}{\sqrt{1 - kV^2}}   \Longrightarrow   L = \sqrt{1 - kV^2}L_0  
\end{equation}
To simultaneously locate the positions of the object's end points, we need a set of synchronized clocks, so the question becomes how to generate such a set. Let us put this question aside for the moment. It will form the subject of section \ref{sec:synchro}.

The second option is to use time dilation. From the generalized Lorentz transformation, the rates of two clocks are related by the formula

\begin{equation}
\label{timedilation}
\Delta t' = \dfrac{\Delta t}{\sqrt{1 - kV^2}}
\end{equation}
But the problem now is to determine how to compare the two readings.

The first thought might be to use a form of the twin paradox. Let us synchronize the clocks of S and S' at a given moment $t = t' = 0$. Let S' run in a very large circle at a constant speed $V$. When the clock in S' eventually returns to face the clock in S, the two elapsed times can be compared. This procedure encounters the same problem as in the original twin paradox, however. Because S' is rotating, it is no longer inertial, the symmetry between the systems is broken and the relativity principle can no longer ensure that both clocks run identically in their own rest frames. 

Of course, accelerated systems can be treated in the framework of relativity-like theories by the use of instantaneously comoving inertial reference frames. But the problem goes beyond purely kinematical effects. Accelerations entail fictive forces and these might influence the clock's mechanism, depending on its specific structure.

These concerns are not hypothetical. The 1971 Hafele-Keating experiment in which cesium clocks were flown around the world and compared with a clock that had remained on the ground is a good showcase for the problems involved in such a method \citep{hafele}. Because the planes were flown at high altitude, Hafele and Keating had to take into account the gravitational influence on the running of the clocks. It turns out that this effect is of a similar magnitude as the kinematic effect due to the plane's velocity. Furthermore, the clock on the ground is not in an inertial system either, because of the daily rotation of the earth, so all clocks must really be compared to an imaginary clock at the center of the earth. The difference in the ticking rate of the clock on the plane and on the ground then depends on the direction of flight of the plane, further complicating the comparison.

The point here is not to analyze the difficulties of the Hafele-Keating experiment, but rather to note how complicated it is to take into account the effects of acceleration, which by the principle of equivalence are of a nature similar to those of gravitational fields. In the case of the Hafele-Keating experiment, the effects had to be calculated out before a proper comparison of the clocks could be performed in order to check just the special relativistic time dilation. 

If we have no previous knowledge of the value of $k$, these calculations run into trouble. For starters, we don't know how the effects of the acceleration compare to the kinematical ones. Second, we will have to introduce many more assumptions. Consider for example Cesium atomic clocks. Their operation depends on the behavior of the hyperfine structure of the atoms, on the interactions of the atoms with magnetic fields (used in the detectors) and on microwave generators used to provide the exciting frequencies. To compare correctly the difference in rates of the clocks S and S', we need to know how each of these components are affected by accelerations (fictive forces). This requires considerably more than just the postulates P1-P3 above, even in gedanken settings. 

In comparison, the second postulate solves the equivalent problem in standard special relativity by offering a simple (gedanken) time-keeping device, namely Einstein's light-clock. This device's proper function as a time-keeping device is ensured by the second postulate itself. But if we reject the postulate, we must replace it by some other assumptions regarding possible clock mechanisms, so that we can compare the workings of two clocks in asymmetrical frames.

The simplest solution to this problem is to sidestep it entirely by keeping both S and S' inertial. The two frames must then move in straight lines and their origins can only coincide once, at which time we synchronize both clocks to be $t = t' = 0$. To compare the ticking rates, let the origin of S' reach the location $x$ in the S-frame at a time $t$ in that frame. At this moment, the clock in $S'$ shows the time

\begin{equation}
\label{timeprimea}
t' = \dfrac{t - vx}{\sqrt{1 - kV^2}}
\end{equation}
Substituting $V = \dfrac{x}{t}$, we obtain that

\begin{equation}
\label{timeprimeb}
t' = \sqrt{t^2 - kx^2}
\end{equation}

In this method there are no accelerations, and the determination of $x$ represents a rest-length measurement in S, which only requires a ruler, so there are no concerns there. But the measurement of the time $t$ now requires two synchronized clocks (one at $x$, the other at the origin). Just like the measurement of $k$ from length contraction, we are back to the need for a synchronization procedure. The Einstein synchronization is based on the second postulate. What are the alternatives?

\section{The conventionality of synchronization}
\label{convent}

As we shall see, the problem of clock synchronization is closely related to the problem of generating a signal that moves at identical speeds in opposite directions. Such a requirement immediately raises the question whether this is connected to the conventionality of synchronization.

The conventionality view posits that we only have empirical validation for the frame invariance of the round-trip speed of light, but that the value of the one-way speed is a matter of convention \citep{reich,grun}. The freedom to choose the value of the one-way speed of light is expressed by Reichenbach's parameter $\epsilon$: denoting the one-way speed of light in the positive x-direction as $\overrightarrow{c}$, we define:

\begin{equation}
\overrightarrow{c} = \dfrac{c}{2\epsilon}
\end{equation}
where $c$ is the round-trip average speed of light, which is a frame invariant quantity. Standard relativity assumes $\epsilon = \dfrac{1}{2}$.

Winnie then derived the so-called $\epsilon$-Lorentz transformations, i.e., the coordinate transformations for arbitrary values of $\epsilon$ \citep{Winnie1, Winnie2}. They explicitly depend on the direction of motion of the system S' with respect to the system S. For example, if S' moves in the positive x-direction, the $\epsilon$-Lorentz transformation for the x coordinate is:

\begin{equation}
x' = \dfrac{x-\overrightarrow{v_\epsilon} t}{\alpha_+}
\end{equation}
where
\begin{equation}
\alpha_+ = \sqrt{\dfrac{\left(c-\overrightarrow{v_\epsilon}\cdot\left( 2\epsilon - 1 \right)\right)^2 - \overrightarrow{v_\epsilon}^2}{c^2}}
\end{equation}

In these equations, $v$ is the speed of the system S' corresponding to the choice $\epsilon = \dfrac{1}{2}$. $\overrightarrow{v_\epsilon}$, on the other hand, is the actual speed of S' as measured by S for a specific but arbitrary choice of $\epsilon$, and it is given by:

\begin{equation}
\overrightarrow{v_\epsilon} = \dfrac{c \cdot v}{c+v(2\epsilon-1)}
\end{equation}

If S' moves in the negative x-direction, however, the transformation takes a different form:
\begin{equation}
x' = \dfrac{x-\overleftarrow{v_\epsilon} t}{\alpha_-}
\end{equation}
where
\begin{equation}
\alpha_- = \sqrt{\dfrac{\left(c+\overleftarrow{v_\epsilon}\cdot\left( 2\epsilon - 1 \right)\right)^2 - \overleftarrow{v_\epsilon}^2}{c^2}}
\end{equation}
and
\begin{equation}
\overleftarrow{v_\epsilon} = \dfrac{c \cdot v}{c-v(2\epsilon-1)}
\end{equation}
Thus, the transformations for a system S' moving in the positive x-direction at speed $\overrightarrow{v_\epsilon}$ differ from the transformations for a system moving at an identical speed in the negative direction (i.e., for $\overleftarrow{v_\epsilon} = \overrightarrow{v_\epsilon}$). The standard Lorentz transformations, on the other hand, correspond to $\epsilon= \dfrac{1}{2}$, for which we have that $\overleftarrow{v_\epsilon} = \overrightarrow{v_\epsilon} = v$ and $\alpha_+ = \alpha_- = \sqrt{1 - \dfrac{v^2}{c^2}}$.

A look at the generalized Lorentz transformations, Eqs.(\ref{final}), shows that they are incompatible with the $\epsilon$-Lorentz transformations for any value of $\epsilon$ other than $\dfrac{1}{2}$. Indeed, suppose that we imagine a body or signal moving in the positive x-direction at speed $\dfrac{1}{\sqrt{k}}$ in some frame S. In any other frame S' moving with respect to S at an arbitrary speed $v$, that same body or signal will move at an identical speed, since, from Eq.(\ref{veladd}),

\begin{equation}
w = \dfrac{\dfrac{1}{\sqrt{k}}-v}{1-kv\cdot\dfrac{1}{\sqrt{k}}} = \dfrac{1}{\sqrt{k}}
\end{equation}

This result is independent of the direction of motion of S' as well as the direction of motion of the signal or body (i.e., a similar result holds for a velocity $-\dfrac{1}{\sqrt{k}}$). The generalized Lorentz transformations therefore imply that the one-way limit speed is identical in all directions. The postulates P1-P3 assume nothing about the properties of such a signal nor even its existence. It might seem surprising, therefore, that they imply a specific choice of $\epsilon$, whether one considers such a choice as conventional or not.

The reason lies in postulate P2 - the (nomological) isotropy of space. For $\epsilon \neq \dfrac{1}{2}$, the $\epsilon$-Lorentz transformations explicitly violate this condition. For example, the length of an object moving at a \textit{given} speed depends on its direction of motion \citep{Winnie1}. Consider a ruler of rest length L' moving at a given speed $\overrightarrow{v_\epsilon} = \eta$ in the positive x-direction, where $\eta$ is some arbitrary constant value. From the point of view of the frame S' its length is:

\begin{equation}
L_+ = L'\sqrt{\dfrac{\left( c - \eta(2\epsilon-1)\right)^2 - \eta^2}{c^2}} = L'\alpha_+
\end{equation}

On the other hand, the same ruler moving at an identical speed (as measured by S) in the negative x-direction (i.e., with $\overleftarrow{v_\epsilon} = \eta$), will have a length given by:

\begin{equation}
L_- = L'\sqrt{\dfrac{\left( c + \eta(2\epsilon-1)\right)^2 - \eta^2}{c^2}} = L'\alpha_-
\end{equation}

Since these lengths are different for $\epsilon \neq \dfrac{1}{2}$, the $\epsilon$-length contraction formula is not invariant under a rotation of the x-axis by $180^0$. This is but one example of explicit isotropy-breaking in the $\epsilon$-Lorentz transformations (time dilation similarly depends on the direction of motion).

To the extent that one considers the choice of $\epsilon$ to be conventional, one will also identify at least part of the postulate of isotropy as conventional. However, this is not the issue here. The generalized Lorentz transformations are only consistent with standard synchronization ($\epsilon = \dfrac{1}{2}$) because they assume full nomological isotropy. Whether such a choice contains an element of conventionality or not is irrelevant. Within the context of one postulate argumentation, isotropy is assumed, and consequently, the one-way maximal speed $\dfrac{1}{\sqrt{k}}$ is isotropic and invariant. The question debated in the present paper is whether we need to know that there is an actual phenomenon that propagates at this speed and what that phenomenon is. The one-postulate position answers in the negative and my claim here is that this will not do. 

We can summarize the argument up to now very simply: the methods of measurement of $k$ that we have seen so far require knowing how to produce a signal that moves identically in two opposite directions. The generalized Lorentz transformations \textit{guarantee} this property in one case \textit{only} - a signal moving at the limit speed $\dfrac{1}{\sqrt{k}}$, and therefore we \textit{must} know which signal possesses this property in order to measure $k$. The only alternative is to add some \textit{other} information, whatever it may be, that allows us to produce such an isotropic signal, at least under some circumstances. But then, this is an additional postulate that must be added to P1-P3 and we are no longer in a so-called one-postulate theory.

The remaining question is whether other methods of measuring $k$ can bypass the problem. As we just saw, using the generalized transformations themselves requires a method of synchronizing clocks. Let us now have a look at that issue, remembering all the while that the postulated isotropy of the transformations restricts us to standard definitions and conventions, i.e., to $\epsilon = \dfrac{1}{2}$.

\section{Synchronizing Clocks}
\label{sec:synchro}

As mentioned in section \ref{intro}, we must perform our analysis as if $k \neq 0 $ even if we don't know this beforehand, because methods adapted to this case will also work for $k = 0$ but the reverse is not true. A case in point is synchronization by clock transport, i.e. synchronizing clocks when they are together and then moving one of them to a new location. If $k=0$ (the Newtonian case), the clocks remain synchronized during transport, but they obviously don't if $k \neq 0$.

Slow-transport synchronization will not solve the problem in this case, because it is merely a formal limit where the transport speed is made to approach zero. It cannot be strictly zero, however, which means that no matter how close we are to the zero transport-speed limit, the clocks will \textit{still} lose their synchronization. Slow transport is a useful tool for some theoretical analyses and rhetorical purposes but it cannot stand as a method of actual synchronization, not even (or rather particularly not) in the gedanken mode. 

It can be very practical, of course, if we only require synchronization up to some finite precision. Then we can calculate how small $kv^2$ should be, i.e., how low $v$ should be, to keep the loss of synchronization within the acceptable bounds. But apart from the fact that this is not what we are after (we need a method of synchronizing clocks in principle, i.e., to absolute precision given ideal technology), in the present context we have a larger problem. Since we don't know the value of $k$ to begin with, we cannot estimate $kv^2$. Thus, there is no way to calculate how slowly we should move the clock in order to keep the de-synchronization below a given threshold. 

If we insisted on using this method nevertheless, we'd have to resort to trial and error: move the clocks, check what time lag is created, repeat more slowly until the time difference is within acceptable bounds. But in this case we wouldn't really need any synchronized clocks at all. For any finite value of the transport speed $v$, we could calculate $k$ from the time difference created and the time dilation formula. But now we are back to the analysis of section \ref{klength}. The problem is again how to compare the readings of the two clocks. A round-trip leaves us with all the problems of asymmetrical accelerations. The other option is, once the first clock has been transported, to move the two clocks back together, at identical speeds in opposite directions until they meet halfway. Provided their motion is identical, the time lag can be ascribed solely to the initial transport. The problem is that once again we require, as in Mermin's method, a mechanism that can move two objects at rigorously identical speeds in opposite directions. But postulates P1-P3 alone are insufficient to provide a method of doing this.

Note, by the way, that if we \textit{did} have such a method, we'd also have a synchronization method that doesn't require slow transport at all. We could use the simpler procedure suggested by N. David \citet[p. 124, note 5]{mermin}: Suppose that we want to synchronize two clocks, one located at the origin and the other at a position $x_2 = L_0$. Place both clocks at the midway position, $x_1 = L_0/2$, and synchronize them. Assuming the two clocks can be moved at identical speeds in opposite directions, transport one clock from the midway position to the origin and the other to the position $x_2 = L_0$. We do not need to know how the clocks move exactly, nor what are their velocities, or indeed whether they are moved uniformly or not. All we need know is that they move identically, which ensures that they remain synchronized. That is the central problem, however.

Synchronization by clock transport fails, therefore, as Mermin's method of measuring $k$ does, because the postulates P1-P3  alone are insufficient to \textit{guarantee} a method of isotropic transport. The remaining option is to synchronize clocks \textit{in situ}, i.e., by some form of Einstein procedure. This requires signaling between the two clocks. But then we need information about such signals.

Suppose for starters that we use material agents. The clocks could be equipped with a gun that fires pellets at a certain velocity, $v_0$. To synchronize two clocks, the observer positions himself midway between the two clocks. The clocks are set so that each fires a pellet at a predetermined time, say when each clock marks the time zero. If the pellets reach the observer simultaneously, the clocks are considered synchronized. If not, the clocks are adjusted and the procedure repeated until they are synchronized. To synchronize a third clock, the observer moves and positions himself midway between clock 2 and 3. In the repeated procedure only clock 3 will be adjusted until it is observed to be synchronized with clock 2. In the same manner any number of clocks can be synchronized with each other.

But this method has the same weakness as the transportation procedure. We must be certain that the pellets are fired at equal speeds. Since the clocks must fire the pellets in opposite directions, however, we cannot determine directly that the pellets move identically. If we build identical guns and rotate one of them, we lose in the process the assurance that nothing has influenced the workings of the firing mechanism; if we build the guns as rotated images of each other, we must still make sure that every possible external influence has also been rotated (this also applies to the case of physical rotations, in addition to possible changes in the internal mechanism). Instead of pellets, we could use some mechanical devices running from the clocks to the observers, but again the same criticism applies. There is no way to be certain that the mechanical messengers run in opposite directions at identical velocities without additional postulates.

Next, we might try non-material signals. That requires significant knowledge about them, however.\footnote{Even determining that the signal is non-material is not trivial. It was not immediately evident to the pioneers of radioactivity that alpha and beta rays were material particles but gamma rays were not.} Suppose for example that we decided to use sound waves to synchronize our clocks. To insure that the signals traveling in opposite directions from the two clocks move at identical speeds, we must know all that might influence the propagation. We must use sealed containers to eliminate air flows, make sure that the temperature, pressure and humidity are uniform and a myriad other adjustments.

We shall quickly face the same problem as with pellets. Unless we can guarantee that every factor potentially influencing the propagation is uniformly distributed or at least invariant under rotation, we have no way of ensuring that our synchronization procedure is sound. Worse, however, is the fact that all this information about sound waves must be added to the theory as additional postulates. Relegating it to the realm of ``empirical evidence" is no different than stating that the constancy of the speed of light is an empirical fact. If we intend to claim that a second postulate is unnecessary, we cannot introduce in its stead highly detailed information about other signals propagation. 

Using light instead of sound makes no difference. We need to know with respect to what ``medium'' or ``system'' the speed of light is identical in both directions and what our velocity is with respect to that ``medium". Of course, Einstein's second postulate is just a way of providing this information, by denying that there is any particular and unique medium with respect to which the velocity is c.  

We see that creating a network of synchronized clocks is a highly non-trivial procedure. Either we assume that $k = 0$ as an additional postulate, or we assume something about signaling methods, whether mechanical or wave-like. In either case, we need additional postulates, and none as simple and elegant as the postulate of constancy of the speed of light.

\section{A weaker second postulate}
\label{sec:weaker}

In the previous sections, we saw repeatedly that a fundamental precondition of experiment is having a signal (material or otherwise) that moves identically in opposite directions. The second postulate ascribes such a property to light, of course, but it apparently requires more - namely that the value of its speed should be invariant in all inertial frames. Is any additional physical information conveyed by this logically stronger requirement? The answer is negative. In addition to the postulates P1-P3, i.e., space-time homogeneity, spatial isotropy and the principle of relativity, one need add only the following postulate in order to derive standard SR:

\textit{Postulate of Light Isotropy}: A light signal propagates with the same speed in all directions, no matter the state of motion of its source.

Although logically weaker than Einstein's second postulate, this assumption does in fact imply that the speed of light is frame-invariant. The proof is very simple. Consider a light source at rest in the frame S, which emits two light signals in opposite directions, say the positive and negative x direction. Let the velocities of the two signals be denoted by $c$ and $-c$ respectively. 

In a frame S' moving with velocity $v$ with respect to S, the signals will appear to move with velocities $c'_+$ and $c'_-$ respectively. The postulate of light isotropy now implies that $c'_- = - c'_+$. These velocities can be calculated from the rule of addition of velocities, Eq.(\ref{veladd}). The requirement  $c'_- = -c'_+$ implies that:

\begin{equation}
\label{iso1}
c'_- = \dfrac{v - c}{1 - kcv} = - c'_+ = - \dfrac{v + c}{1 + kcv}   ;    
\end{equation}

From this, we immediately obtain:

\begin{equation}
k = \dfrac{1}{c^2}
\end{equation}
which also immediately implies that $c'_+ = c$. Thus any signal that verifies the postulate of isotropic propagation must propagate at the invariant speed $k^{-1/2}$.

The present analysis directly contradicts, therefore, Mermin's assertion, already quoted in section \ref{intro}, that

\begin{quote}
It is not, however, necessary for there to be phenomena propagating at the invariant speed to reveal the value of k.\citep[p. 123]{mermin}
\end{quote} 

On the contrary, measuring $k$ (as well as, e.g., creating a network of synchronized clocks) require an isotropically-propagating signal and this property is only guaranteed if the signal propagates at speed $k^{-1/2}$. \textit{A fortiori}, we require that such a signal exists. Were it not the case, the theory might still be meaningful, of course, but at the expense of additional postulates.

\section{Musings about the function of the second postulate}
\label{function}

Recall L\'{e}vy-Leblond's opinion, quoted in section \ref{intro}, that the properties of light are over-emphasized in SR, an opinion later echoed by Mermin and others. Yet we have seen that something goes awry when we pry into the physical meaning of the transformations and we hit a snag every time we try to measure $k$. In fact, the situation is worse for not only the determination of $k$ seems to require additional postulates, but practically all measurements are in trouble (in particular, it becomes much more difficult to ensure that different observers use identical units of measurements \citep{drory2}).

Why should that be? Is there any fundamental reason why we need specific information about the propagation of some signal? I do \textit{not} claim that SR must be based on light, as opposed to some other signal, and in this sense (but only in this sense) I agree that SR is not necessarily related to electromagnetism. Other postulates could work equally well to establish the prerequisites of space-time measurements. But neither can special relativity be said to follow from space-time symmetries and the principle of relativity alone. 

At its deepest level, special relativity can be thought of as a theory that unifies space and time into space-time with a four-dimensional metric. Thus, special relativity revises our understanding of the geometry of space and time, which is essentially the relation between different points in space-time. But an observer is always a spatially local entity, in fact a point-like entity in three dimensions, or equivalently, a one-dimensional world line in space-time. In the end, all observations and  measurements must reach this local entity for analysis. 

There is therefore a fundamental gap between the referent of the theory - the four-dimensional space-time continuum - and the one-dimensional world-line of the observer. In order for the space-time measurements to be properly compared and related, such as in the generalized Lorentz transformations, we need some method of transferring information from different space points to the single location of the observer. This is fundamental to the meaning of special relativity as a theory of space-time structure. Transporting clocks, shooting pellets, signaling, are all methods of achieving this. Determining the value of $k$ requires such transport of information; without the value of $k$ the transformations are useless. It is the second postulate that allows us to analyze such information transmission. Without it, or some equally specific postulate on signaling, the theory is literally blind. Transfer of information is thus a necessary precondition of any experiment and is required, in general, to make physical sense of the theory in a way that Newtonian physics does not require.

\section{Summary}

The content of innovative physical theories lies not just in new equations. They often involve redefining previously established concepts, revising fundamental assumptions and at their best, reforging our world-view. Once such a theory has been accepted by the scientific community and become established, it can be easy to view these other changes as independent of the theory. At this point, there is often a tendency to axiomatize the theory and to seek its simplest possible mathematical formulation. There is much to be gained from such attempts, but in the process, we are facing a danger of over-formalization. 

Special relativity is almost unique in being a theory that emerges from a revision of the preconditions of experiments. It is intimately linked to the question of how we measure time and space coordinates and what the necessary preconditions for such measurements are. Such analysis is so much a part of our world-view nowadays that we may think it can be disconnected from other aspects of the theory. I venture no opinion whether this is the case or not, but I do argue that the importance of the postulate of the constancy of the speed of light does not lie only (or even mostly) in the formal completion of the equations. The postulate is much more than merely a claim regarding the value of some free parameter in the theory. Instead, it plays an important part in the analysis of the preconditions of experiments that form the core of special relativity.

In this paper I have analyzed whether we could empirically determine the value of the parameter $k$. Direct use of the transformations to extract $k$ from length contraction or time dilation fail because they require a set of synchronized clocks as a prerequisite to the measurement. Procedures for synchronizing clocks must rely on either signaling or mechanical transportation (either of the clocks themselves or of some material carriers of information). Both these methods fail if further assumptions are not added. These assumptions may be dynamical (in the case of mechanical transportation) or kinematical (about the velocity of non-material signals). In either case, they amount to a second postulate of some kind. Among possible postulates, it appears to me that the Einstein postulate is particularly elegant and simple, though this may be a personal preference. 

Methods that do not require synchronized clocks still fail without additional postulates. These could in turn be used to create a synchronized network of clocks. Mermin's method of unit runners, for example, requires ensuring that some device moves identically to the left and to the right, a construct that would allow clock synchronization. I have argued that we have no way to ensure such symmetry without additional assumptions, however. Precisely the same problems plague attempts to use the Doppler effect to determine the value of $k$.

The required information is logically weaker than Einstein's assumption of constancy of the speed of light. This is the postulate of isotropy of light speed, which states that the velocity of light is always identical in all directions, no matter the state of the emitter. By adding this postulate to the principle of relativity and the space-time symmetries, one recovers the Lorentz transformations together with their physical underpinnings. 

This shows that, given the postulate of isotropy - which eliminates the freedom of choice of $\epsilon$ - the core of the second postulate lies in providing us with a phenomenon that propagates with identical speeds in opposite directions. This function, however trivial, contingent or specific it might seem, turns out to be crucial to the physical content of the theory.

\end{document}